\documentstyle[11pt,newpasp,twoside,epsfig]{article}

\begin{document}

\title{Effect of the large scale environment on the internal dynamics
of early-type galaxies}
\author{Maubon, G. \& Prugniel, Ph}
\affil{CRAL - Observatoire de Lyon, France}

\begin{abstract}
We have studied the population-density relation in very sparse
environments, from poor clusters to isolated galaxies, and we find
that early-type galaxies with a young stellar population are
preferably found in the lowest density environments. We show a marginal
indication that this effect is due to an enhancement of the stellar
formation independent of the morphological segregation, but we failed
to find any effect from the internal dynamics.
\end{abstract}

\keywords{Galaxies, Evolution, Population, Environment}

\section{Background}
It is generally accepted that the rate of star formation in early-type
galaxies is enhanced in low-density environments (Schweizer \& Seitzer
1992, de Carvalho \& Djorgovski 1992, Guzm\'an et al. 1992, Rose et
al. 1994, J{\o}rgensen \& J{\o}nch-S{\o}rensen 1998, Bernardi et
al. 1998).  It results in a population - density relation: The
metallic features in the spectra (eg. Mg2) are weaker and the Balmer
lines stronger in low-density regions.

Recently (Prugniel et al. 1999) we have analyzed the
population-density relation in low-density environments (isolated
galaxies to poor clusters).  We found that the early-type galaxies
which are likely to contain a young sub-population are mostly found in
the sparsest environments.

The approach used to diagnostic this effect is to study the
Mg$_{2}$--$\sigma_{0}$ and the Fundamental Plane (FP)
relations. Indeed, although both relations are sensitive to the
dynamics and to the stellar content, departures resulting from one or
the other origin will have opposite signs. One the one hand, if a
galaxy is found below the Mg$_{2}$--$\sigma_{0}$ relation (see
Fig. 1), it may have either an unusually low velocity dispersion or a
high Mg$_{2}$ index. On the other hand, a galaxy below the FP relation
has either a low velocity dispersion or contains a young population.

\begin{figure} \plottwo{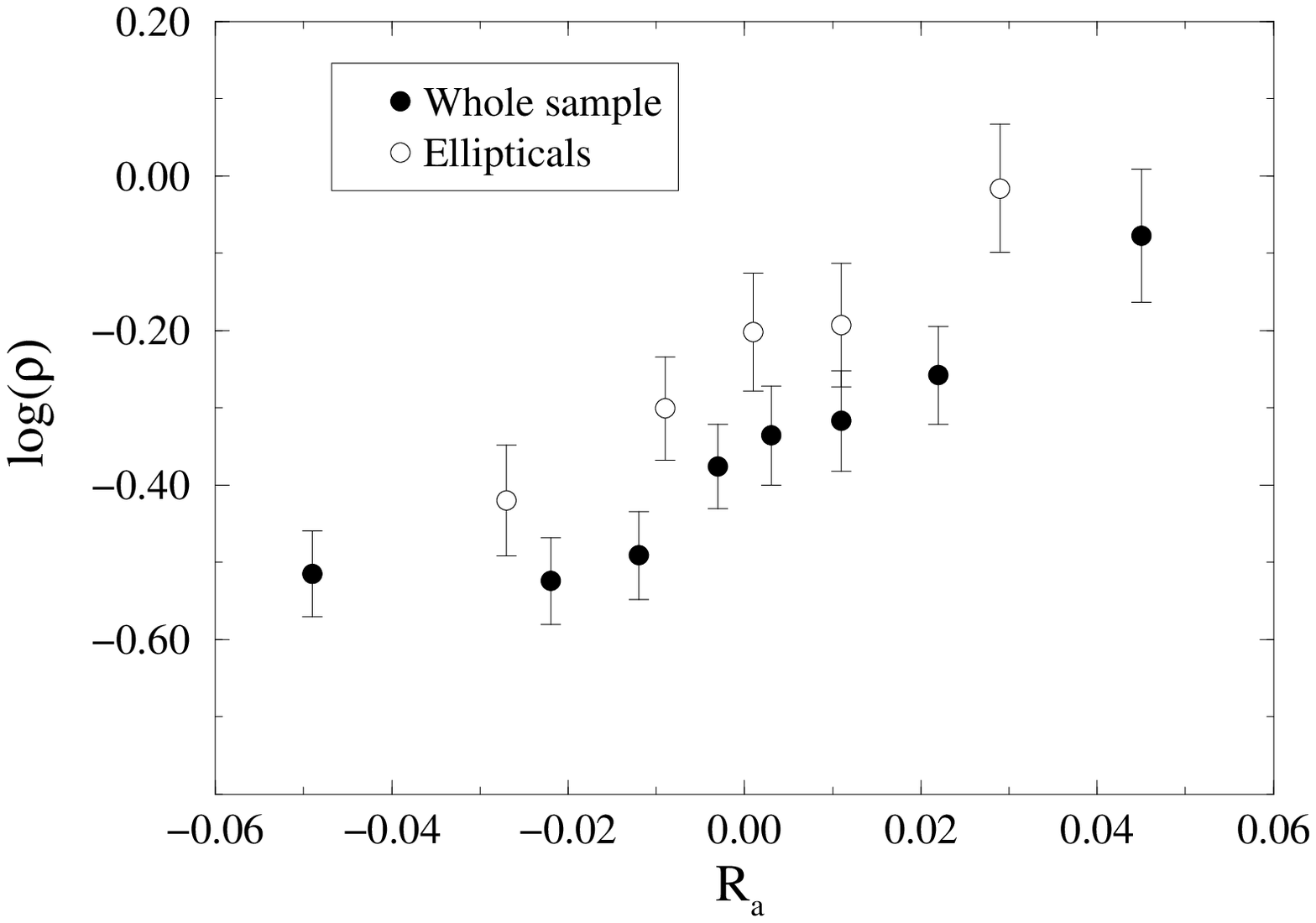}{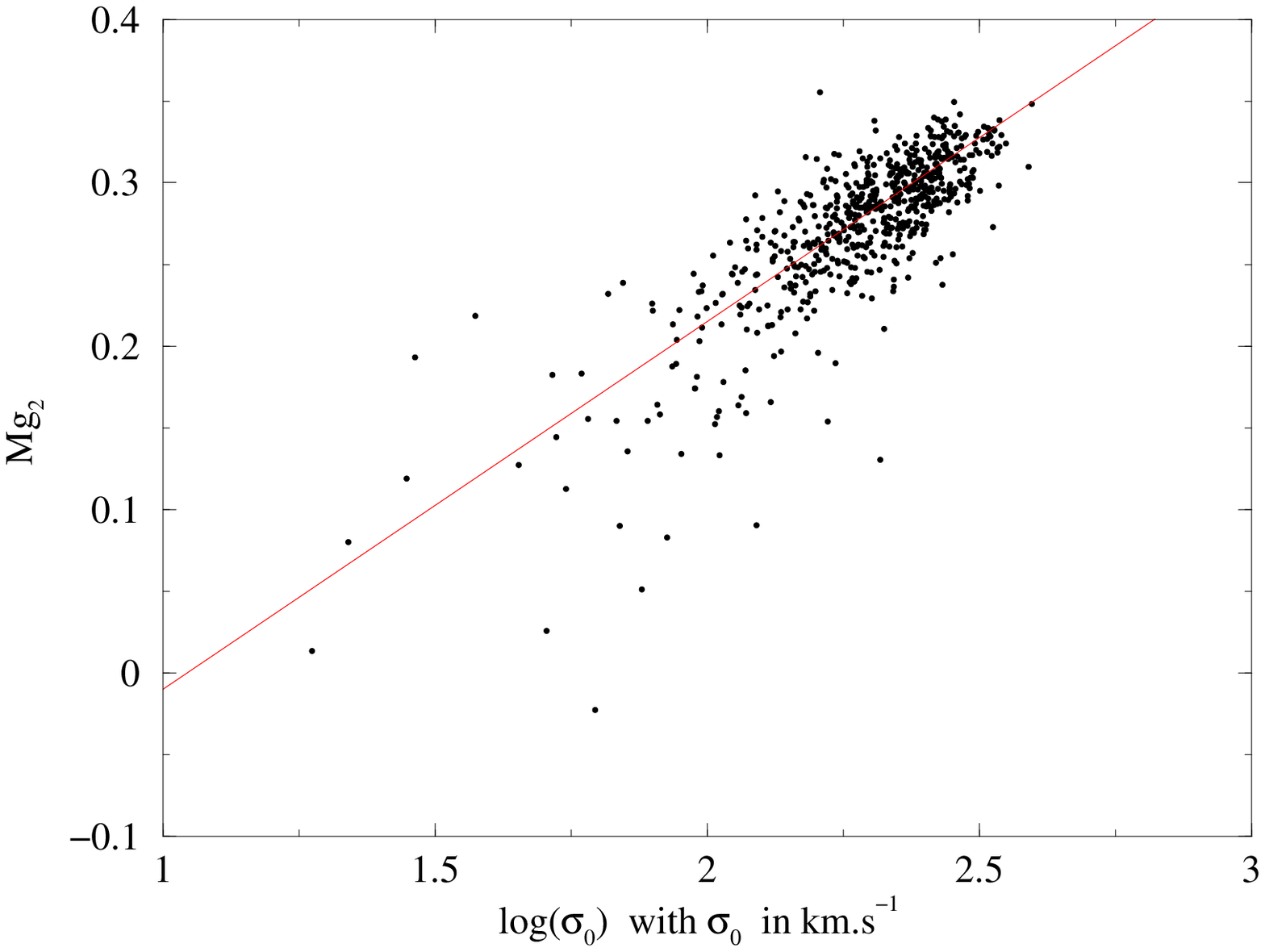}
\caption{Left : Mean density of the environment ($\log{\rho}$) vs
residuals of the Mg$_{2}$-$\sigma_{0}$ relation for the whole sample
(filled symbols, 80 objects in a point) and \emph{bona-fied}
ellipticals only (open symbols, 40 objects in a point) The error bars
are 1$\sigma$ uncertainties. Right: The Mg$_{2}$--$\sigma_{0}$ diagram
of our sample with the classical relation
Mg2=0.225*$log(\sigma_{0})$-0.235.}.
\end{figure}

Our analysis of the residuals shows that the environment has primarily
a visible effect on the stellar content (not on the dynamics).

\section{Dynamical evolution}
It is not a surprise that an enhancement of the young population is
detected: A relatively small fraction of young stars is sufficient to
significantly modify the broad-band colors and the line strength
indices.

However, the enhancement of the stellar formation should have a long
term effect on the dynamics of the galaxies. This delayed star
formation in early-type galaxies is mostly occurring in the central
regions and hence modifies the mass balance in these galaxies.

Unfortunately, when we into account the stellar population effect as
deduced from the residuals to the Mg2-$\sigma$ relation, no more
environmental effect persist in the FP analysis. We cannot find any
dynamical evolution with this analysis. However, the FP analysis
mostly diagnoses the equilibrium status of the galaxies, and this
long term evolution is not expected to significantly disturb the
equilibrium.  However, the FP analysis is also sensitive to the
details of the dynamics and of the structure (see Prugniel et
al. 1997).

Using data collected in the Hypercat database, in particular the
kinematic and photometric profiles, we are trying to
study the systematics of these non-homologies and their relation with
the environment.


\begin{references}
\reference Bernardi, M., Renzini, A., da Costa, L.N. et al., 1998,
\apj\ 508, L143 \reference De Carvalho, R., R., Djorgovski, S., 1992,
\apj\ 389, L49 \reference Guzm\'an, R., Lucey, J.\,R., Carter, D.,
Terlevich, R.\,J., 1992, \mnras\ 257, 187 \reference J{\o}rgensen, I.,
J{\o}nch-S{\o}rensen, H., 1998, \mnras\ 297, 968 \reference Prugniel,
Ph., Simien, F., 1997, \aap\ 321, 111 \reference Prugniel, Ph., Golev,
V., Maubon, G., 1999, \aap\ 346, L25 \reference Rose, J., Bower, R.,
Caldwell, N. et al., 1994, \aj\ 108, 2054 \reference Schweizer, F.,
1982, \apj\ 252, 455 \reference Schweizer, F., Seitzer, P., 1992, \aj\
104, 1039
\end{references}
\end{document}